\documentclass[aps,prd,letterpaper,twocolumn,nofootinbib,showpacs,preprintnumbers]{revtex4-1}

\usepackage{amsmath,amsfonts,amssymb}
\usepackage{bbm}
\usepackage{xcolor}
\usepackage{graphicx}
\usepackage[bookmarks,bookmarksnumbered]{hyperref}

\newcommand{\be}{\begin{equation}}
\newcommand{\ee}{\end{equation}}
\newcommand{\bea}{\begin{eqnarray}}
\newcommand{\eea}{\end{eqnarray}}
\newcommand{\besub}{\begin{subequations}}
\newcommand{\eesub}{\end{subequations}}
\newcommand{\ba}{\begin{array}}
\newcommand{\ea}{\end{array}}
\newcommand{\bi}{\begin{itemize}}
\newcommand{\ei}{\end{itemize}}
\newcommand{\nn}{\nonumber}

\newcommand{\vev}[1]{\ensuremath{\langle #1 \rangle}}
\newcommand{\GeV}{{\rm GeV}}
\newcommand{\TeV}{{\rm TeV}}

\newcommand{\Ocal}{{\cal O}}
\newcommand{\mgr}{\ensuremath{m_{3/2}}}

\newcommand{\lf}{{16 \pi^2}}

\newcommand{\tb}{{\tan \! \beta}}

\begin{document}

\preprint{DO-TH 11/01}
\title{Flavorful hybrid anomaly-gravity mediation}
\author{Christian Gross}
\author{Gudrun Hiller}
\affiliation{Institut f\"ur Physik, Technische Universit\"at Dortmund, D-44221 Dortmund, Germany}
\begin{abstract}
We consider supersymmetric models where anomaly and gravity mediation give comparable contributions to the soft terms and discuss how this can be realized in a five-dimensional brane world. 
The gaugino mass pattern of anomaly mediation is preserved in such a hybrid setup.
The flavorful gravity-mediated contribution cures the tachyonic slepton problem of anomaly mediation. 
The supersymmetric flavor puzzle is solved by alignment.
We explicitly show how a working flavor-tachyon link can be realized with Abelian flavor symmetries and give the characteristic signatures of the framework, including ${\cal{O}}(1)$ slepton mass splittings between different generations and between doublets and singlets.
This provides opportunities for same flavor dilepton edge measurements  with missing energy at the Large Hadron Collider (LHC).
Rare lepton decay rates could be close to their current experimental limit.
Compared to pure gravity mediation, the hybrid model is advantageous because it features a heavy gravitino which can avoid the cosmological gravitino problem of gravity-mediated models combined with leptogenesis.
\end{abstract}
\maketitle

\section{Introduction}

Weak scale supersymmetry (SUSY) is an attractive framework for physics beyond the standard model (SM) since it can stabilize the weak scale and provide a candidate for dark matter.
SUSY also faces new challenges, in particular how to avoid excessive flavor- and \emph{CP} violation:
While in the SM flavor changing neutral currents (FCNCs) and \emph{CP} violation are controlled by small mixing angles and mass splittings, mechanisms of similar power need to be built
into TeV-scale models to pass the experimental constraints from flavor physics (see {\it e.g.}~\cite{Isidori:2010kg} for a recent review).

Anomaly-mediated supersymmetry breaking (AMSB) \cite{Randall:1998uk,Giudice:1998xp} is a model that fulfills these selection criteria
since it features flavor violation induced by the SM Yukawas only, {\it i.e.,}  it is minimally flavor violating. Moreover, in the limit where the top Yukawa flows to its infrared fixed point, AMSB becomes  approximately flavor-blind~\cite{Allanach:2009ne}.
On the other hand, pure AMSB leads to tachyonic sleptons when the visible sector is the minimal supersymmetric SM (MSSM).
There are many proposals to cure this problem (see {\it e.g.}~\cite{Randall:1998uk,Pomarol:1999ie,Gherghetta:1999sw,Gregoire:2005jr}); none of them is agreed upon to be completely satisfactory in all regards, however.
At the same time, an old challenge inherited from the SM remains:
While  the spectra and mixings  of SM matter are described by the SMs Yukawa matrices, the origin of their peculiar texture is not addressed and without further ingredients not accessible.

Here, we consider a setup where the tachyon problem is solved by order one flavorful contributions to the soft terms from Planck-scale-mediated (a.k.a.~gravity-mediated) SUSY breaking (PMSB).
Usually, when studying AMSB, it is assumed that the PMSB contribution vanishes (or is subleading), {\it i.e.,} that the SUSY breaking sector is sequestered.
Here, we require that PMSB is of comparable size as AMSB.
We term the SUSY breaking sector {\it semi-sequestered} in this case.
In Sect.~\ref{sec:hybrid} we discuss  how such a situation could be realized.

Since the PMSB contribution generically carries flavor violation beyond the SM, hybrid  anomaly-gravity mediation is  a non-minimally flavor violating model.
As such, if evidenced, it can in principle lead to a theory of flavor by distinguishing between different mechanisms for  generating the observed family structure (see, for instance,~\cite{Feng:2007ke,Nomura:2007ap,Hiller:2010dv}).

In the hybrid setup, the large order one contributions from PMSB flavor are tamed by sufficiently aligning SM fermions and sfermions~\cite{Nir:1993mx}, resulting in a  viable flavor phenomenology. 
Unlike models with a flavor-blind solution to the tachyonic slepton problem, this also leads to interesting signals in the flavor sector.

Besides addressing the tachyon  together with the flavor problem, hybrid anomaly-gravity mediation has the virtue of a heavy gravitino with mass  $\mgr \sim {\cal{O}}(10-100)\ \TeV$. This is beneficial
for a sensible cosmology when baryogenesis proceeds via leptogenesis, which is a very elegant baryogenesis method in case that the neutrino masses are generated by a seesaw mechanism.
In this regard, the hybrid model improves on pure gravity mediation, which has a gravitino mass of the order of the superpartner masses.

In Sect.~\ref{sec:hybrid} we discuss how comparable anomaly-mediated and gravity-mediated soft terms could arise in a 5-dimensional (5d) brane world.
In Sect.~\ref{sec:flavormodel}, we work out the low energy spectrum and illustrate how both the observed matter flavor structure and the required lepton-slepton alignment can be explained by a simple Froggatt-Nielsen (FN) 
flavor symmetry. 
We give predictions for the LHC, and low energy precision experiments with or without \emph{CP} violation, in Sect.~\ref{sec:pred} and conclude in Sect.~\ref{sec:conclusions}.

\section{Semi-sequestering} \label{sec:hybrid}

We briefly recall the mechanism of anomaly mediation:
AMSB is usually derived by formulating supergravity in the superconformal tensor calculus where a Poincar\'e-invariant  vacuum expectation value (VEV) $\vev{\Phi} = (1,0,F_{\Phi}) $ of the superconformal compensator superfield $\Phi$ breaks superconformal symmetry to super-Poincar\'e symmetry.
Couplings of $\Phi$ to the visible sector induce visible sector SUSY breaking when $F_{\Phi}\neq 0$.
These couplings are determined by superconformal invariance.
The MSSM is classically invariant (in the absence of a $\mu$-term), which implies that $\Phi$ does not couple to visible sector superfields and no SUSY breaking is induced by $F_{\Phi}$. 
However, the symmetry is anomalous (hence the name) so that $\Phi$ {does} couple to the visible sector at quantum level.
It turns out that this induces all soft terms~\cite{Randall:1998uk,Giudice:1998xp}.
The gaugino masses $M_{\lambda}$, the trilinear scalar couplings $A$ and the soft scalar mass
terms $m^2$ are given by
\besub \label{eq:am}
\begin{alignat}{3}
M_{\lambda} & =  F_{\Phi} \beta_{g}/g && \sim \pm   M_{\Phi} g^2\,,\label{eq:ama}
\\
A& =  -F_{\Phi}\beta_{Y}  &&\sim M_{\Phi}\ Y(Y^2-g^2)\,,\label{eq:amb}
\\
m^2& =  \frac{1}{2}|F_{\Phi}|^2\mu\frac{d}{d\mu}\gamma &&\sim |M_{\Phi}|^2 (\mp g^4-g^2 Y^2 +Y^4)\,, \label{eq:amc}
\end{alignat}
\eesub
where $\gamma$ is the chiral superfield anomalous dimension, $\beta_{g}$, $\beta_Y$ are the $\beta$-functions for the gauge- and Yukawa couplings, respectively, and we defined $M_{\Phi}=F_{\Phi}/\lf$. 
Note that $F_{\Phi} \simeq \mgr$.
The expressions after the $\sim$ symbol are to be understood symbolically, {\it i.e.},~numbers and indices are omitted.
The signs depend on the sign of $\beta_{g}$.

As discussed in~\cite{Randall:1998uk}, the soft terms (\ref{eq:am}) generically are negligible compared to PMSB, because the cancellation of the vacuum energy typically requires $F_{\Phi} \sim F_S/M_P$. Here, $S$ is the hidden sector superfield with the highest $F$-term VEV, $F_S$,
and $M_P$ denotes the Planck mass.
$M_*$-suppressed higher-dimensional operators (where $M_*$ is a high mass scale such as the Planck- or string scale), which arise from integrating out unknown physics at $M_*$, generically give a contribution to the soft masses of the order $m^2 \sim (F_S/M_*)^2$, which is larger than the AMSB contribution by an inverse loop factor times $(M_P/M_*)^2$.

Here, we would like to realize the situation where the scale of the PMSB-induced soft terms is suppressed with respect to its natural size $F_S/M_*$, but only so much that it is still comparable to AMSB.
As argued in the Introduction, this is motivated on one hand by the desire to have a gravitino mass which is $\Ocal(100)$ larger than the soft masses ($\to$~PMSB should not be dominant) and on the other hand by requiring that the PMSB contribution to the slepton masses cures the tachyonic slepton problem ($\to$~PMSB should not be subleading).

The suppression of PMSB with respect to its natural scale can be justified in 5d brane models where the MSSM sector and the SUSY breaking sector are located on different branes~\cite{Randall:1998uk}. (An alternative is {conformal sequestering}~\cite{Luty:2001jh} in 4d models by strong and conformally invariant dynamics.)
Just as in 4d hidden sector models, there are no direct renormalizable couplings between the two sectors.
Here, the SUSY breaking sector is more than hidden however:
Because of locality, the effective visible-hidden sector couplings which are induced by the exchange of heavy bulk fields are exponentially suppressed.
Before proceeding, we should mention that, with more than one extra dimension, spatial separation of the hidden and visible sectors is not a sufficient condition for this suppression (see {\it e.g.}~\cite{Anisimov:2001zz}).
For that reason we stick to 5d brane models.

The MSSM gauge multiplets could live either in the bulk or on the visible brane.
In the former case, SUSY breaking can proceed by the mechanism of gaugino mediation~\cite{Kaplan:1999ac}, or, in the absence of a hidden sector singlet, by radion mediation~\cite{Chacko:2000fn}. 
Hybrid radion-anomaly mediation is a particular case of mirage mediation. 
Such a SUSY breaking scheme is naturally realized in models based on flux compactifications and has been extensively studied~\cite{Choi:2004sx}.

Here, we consider the case where both the MSSM gauge and matter fields are located on the visible brane, and the only light fields in the bulk come from minimal 5d supergravity.
In this case, the following effects can contribute to the soft terms:
(a)~AMSB (Eq.~(\ref{eq:am})), (b)~the aforementioned exponentially suppressed, but not necessarily negligible effects from the exchange of massive bulk fields and, (c), loop effects of massless 4d supergravity (SUGRA) modes and the radion.
The latter are subleading if the distance between the branes is large enough, and we assume that this is the case.\footnote{On the other hand, it was proposed~\cite{Randall:1998uk} that it is just these loop corrections that cure the tachyonic slepton problem, assuming that the branes are close enough to each other.
While explicit calculations showed that the contributions to the soft masses due to hidden sector $F$-terms are negative and thus worsen the slepton problem~\cite{Buchbinder:2003qu},
it turns out that $D$-terms give rise to positive soft masses~\cite{Gregoire:2005jr}.
}
Note also that the radion $F$-term and the Kaluza-Klein modes of the 5d SUGRA multiplet do not mediate SUSY breaking at tree-level in this setup~\cite{Luty:1999cz}.

What we mean by {\it hybrid anomaly-gravity mediation} is a combination of (a) and (b).
Let us discuss the effect (b) in more detail.
Bulk fields with a mass $M_*$ induce, by the exchange of a single propagator, effective couplings which are suppressed by $e^{- M_* L}$, where $L$ is the distance between the branes.
(This follows from the fact that a position-space propagator of a field with mass~$m$ linking two points separated by a distance~$d$ is suppressed by $e^{- m d}$.)
This induces, among others, the operator 
\be \label{operator}
\frac{e^{-M_* L}}{M_*^2} \ X_{ij} \ S \bar S Q_i \bar Q_j  \big|_{\theta^4}
\ee
in the 4d effective Lagrangian.
Here, $X$ is a Hermitian matrix with, in the absence of a flavor model, order one entries, and $Q_i$ is a matter superfield with family-index~$i$.
The operator (\ref{operator}) leads to the following PMSB contribution to the soft scalar mass terms:
\be \label{PMSBsoft}
m^2|_{\rm PMSB}
\sim 
\frac{e^{-M_* L}}{M_*^2} \ |F_S|^2  \ X
\sim
|M_{\Phi}|^2\ r X \,,
\ee
where we defined
\be \label{r-parameter}
r
=
(\lf M_P/M_*)^2 e^{-M_* L}
\ee
and we used $F_S/M_P \sim F_{\Phi}$ in the last relation.
For only moderately large values of $M_* L$, $r$ is tiny and the PMSB contribution is negligible compared to AMSB, as usually assumed.
Here, by contrast, we assume that $r$ is of order unity so that $m^2|_{\rm PMSB}$ is comparable to 
the AMSB contribution to the sfermion masses given in Eq.~(\ref{eq:amc}).
This requires $e^{-M_* L}$ to be approximately equal to the first factor in (\ref{r-parameter}), {\it i.e.}, one needs $M_* L\simeq10+\ln(M_P^2/M_*^2)$, which involves some tuning.\footnote{We do not deal with the question if and how this scenario could be embedded in string- or M-theory.
An interesting starting point for such a study seems to be heterotic M-theory~\cite{Witten:1996mz}, which is a natural candidate for the UV completion of 5d brane models.
However, a problematic aspect could be the fact that the bulk of heterotic M-theory derived 5d brane models contains, apart from 5d supergravity modes, many additional moduli.
These would generically spoil our setup if they are lighter than $L^{-1}$.
The situation is maybe not hopeless, however: It was argued~\cite{Kachru:2006em} that under certain conditions a part of the moduli can obtain masses which are parametrically larger than $L^{-1}$, while others obtain masses which scale like $L^{-1}$.}

Note that an analogous operator as~(\ref{operator}) leads to a contribution of the order $r |M_{\Phi}|^2$ to the $B \mu$-term.
On the other hand, no $\mu$-term, gaugino masses and $A$-terms are induced by PMSB when $S$ is not a gauge singlet~-- which we assume.
Then (i) the distinctive pattern of gaugino masses of pure AMSB is maintained and (ii) the $\mu$-problem of AMSB persists~-- we assume that a phenomenologically viable $\mu$-term is generated by an unspecified mechanism.

\section{Aligning flavor} \label{sec:flavormodel}

In this section, we discuss the expected slepton spectrum for the case that the SUSY flavor problem is solved by alignment only.
We also propose explicit examples which realize sufficient alignment and realistic lepton masses and mixings with $U(1) \times U(1)$ FN models.

\subsection{Slepton mass terms}

We start with the slepton soft terms at the high scale (near $M_*$).
The slepton mass squared matrix in hybrid anomaly-gravity mediation reads
\be \label{eq:init}
{\mathcal M}^{2}_{M} 
\sim 
|M_{\Phi}|^2 (- g_M \mathbf{1} + r  X^M ) \, ,
 \ee
where the first term stems from anomaly mediation, while the second one denotes the PMSB contribution. 
The $X^M$  (cf.~Eq.~(\ref{operator})) are hermitian matrices in flavor space.
Here, we defined $g_{L}:=  (99/50) g_1^4 + (3/2) g_2^4$ and $g_{R}:=(198/25) g_1^4$ in terms of the $SU(2)_L$, $U(1)_Y$ gauge couplings $g_2, g_1$, respectively.
The label $M=L (R)$ refers to $SU(2)_L$ doublets (singlets). 
In Eq.~(\ref{eq:init}), we omit for brevity  the flavor non-universal AMSB part proportional to powers of lepton Yukawas, see Eq.~(\ref{eq:amc}). For moderate values of $\tb$, to which we restrict ourselves in this work, they  are all $\ll1$ and their contribution to the slepton mass is subleading to the flavor nonuniversal PMSB one. 

Furthermore,  the following approximations  have been made in Eq.~(\ref{eq:init}):
(i) 
The $F$-terms are neglected, because these are suppressed by $(m_{l_{i}}/\tilde m)^2 $,
where $m_{l_{i}} $ denotes the $i$th-generation charged lepton mass  and $\tilde m$ 
the average slepton mass scale.
(ii) 
The $D$-terms have been dropped, because these are suppressed with respect to the soft terms
by $(m_Z/\tilde m)^2$, where $m_Z$ is the mass of the $Z$ boson and stands for the weak scale in the remainder of this work.
(iii)
We neglect the chirality mixing blocks, because these are suppressed with respect to the chirality preserving blocks. Specifically, the flavor diagonal elements are suppressed by 
$\tb \ m_{l_{i}}/\tilde m$.

\subsection{Spectrum}

To compare our model with data, we need to evolve the slepton mass terms to the weak scale.
The MSSM renormalization group (RG) evolution dampens both flavor mixing and the tachyon 
problem by inducing an order one, positive flavor blind contribution to ${\mathcal M}^{2}_{M} $ from the gaugino masses.  
Since the latter stem from AMSB alone in our semi-sequestered model, and we consider flavor models which are accurate up to numbers of order one only, we can safely approximate the slepton mass matrices at the weak scale by Eq.~(\ref{eq:init}) with $g_M$ taken at the weak scale, numerically $g_{L,R} \simeq 0.3$. 
Hence, the parameter $r$  must be ${\cal{O}}(1)$ for both doublet and singlet sleptons to get rid of tachyons. 

Note that in Eq.~(\ref{eq:init}) we followed Eq.~(\ref{operator}) and normalized each diagonal element as $X_{ii}^M \sim 1 $. We also require all sleptons to have masses linked to the electroweak scale. 
The latter condition could be relaxed for
 the sfermions of the first and second generation, which could be heavier, {\it i.e.,} $X^M_{ii} \gtrsim 1$ for  $i=1,2$, but we do not entertain this possibility here.

The PMSB contributions to the soft terms given in Eq.~(\ref{eq:init})
are not expected to be flavor-blind because in general they
pick up flavor breaking from physics at $M_*$ and above. 
Therefore, the mass splittings between sleptons of different generations,  $\Delta \tilde m^2_{Mij} =\tilde m_{Mi}^2 - \tilde m_{Mj}^2 $, as well as those between the sleptons within the same generation but with different $SU(2)_L$ transformation, $\Delta \tilde m^2_{LRi} = \tilde m_{Li}^2 - \tilde m_{Ri}^2 $, are all large:
\besub \label{eq:masssplit}
\bea   
\label{eq:splitij}
\Delta \tilde m_{M ij}^2 &\sim& r  |M_{\Phi}|^2\,,
\\
\label{eq:splitLR}
\Delta \tilde m_{LRi}^2 &\sim& r  |M_{\Phi}|^2 \,.
\eea
\eesub

\begin{figure}
 \centering
\includegraphics[width=0.5\textwidth]{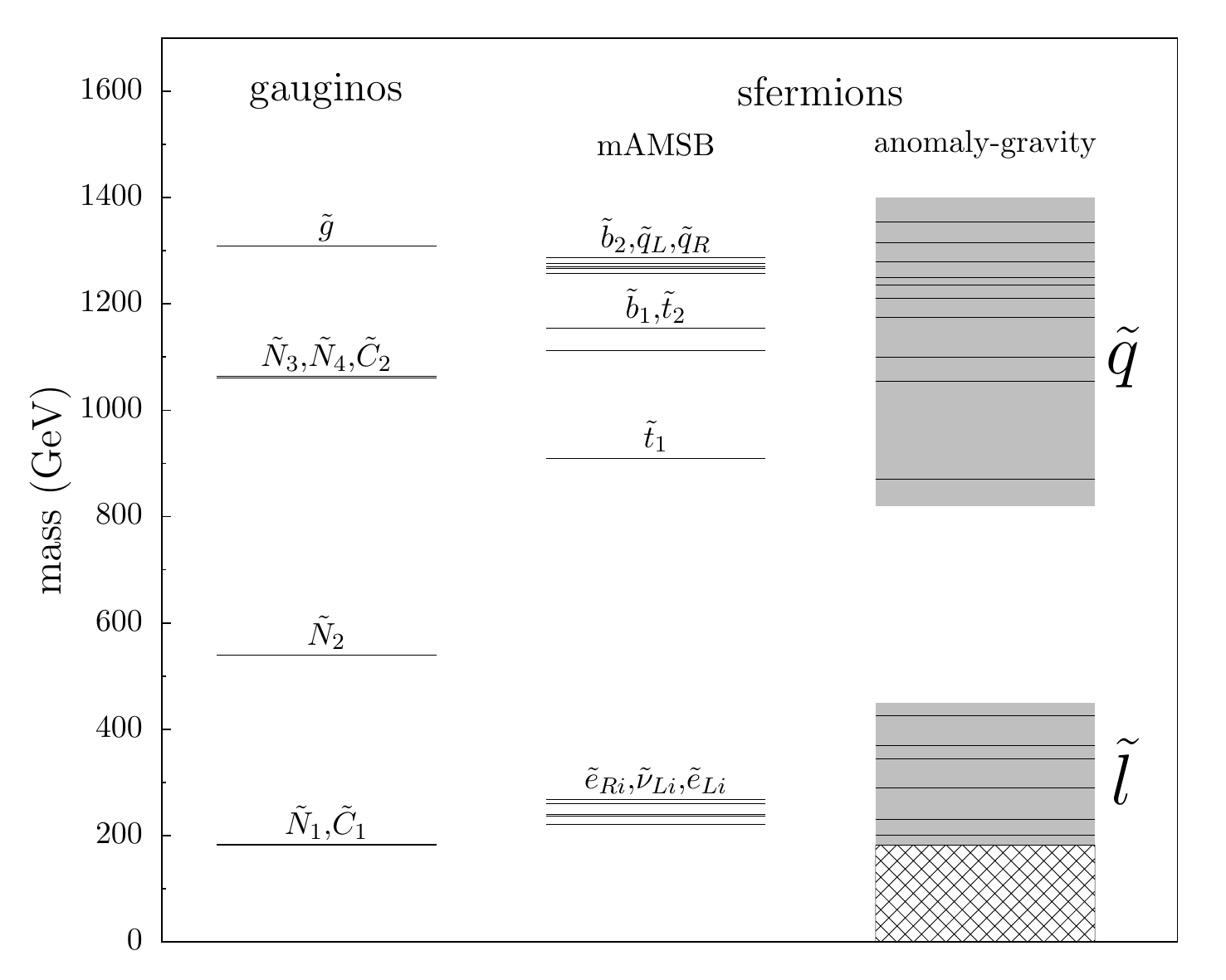}
\caption{Schematic plot of a sample sfermion spectrum in hybrid AMSB-PMSB (3rd column), compared to an mAMSB spectrum (2nd column) with universal scalar mass uplift $m_0= 350 \ \GeV$ and $\tb =5$, $m_{3/2}=60 \ \TeV$, $\mu >0$.
The gaugino masses (1st column) are as in pure AMSB (and mAMSB).
The crosshatched band is disfavored phenomenologically.
}
\label{fig:spectra}
\end{figure}

As an illustration we show a sample superpartner spectrum comparing minimal AMSB (mAMSB) with a
universal slepton mass contribution $m_0$ to hybrid anomaly-gravity mediation
in Fig.~\ref{fig:spectra}.  
The plot is made for $m_0= 350 \ \GeV$, $\tb =5$, $m_{3/2}=60 \ \TeV$ and a positive $\mu$-term.
We use the ISAJET spectrum generator~\cite{Paige:2003mg} for the gaugino and the mAMSB sfermion masses. 
The gaugino masses (1st column) of the hybrid model are as in pure AMSB with the mass terms in the soft breaking Lagrangian respecting the hierarchy
$|M_1| : |M_2| : |M_3|$ of $3:1:7$, featuring an almost degenerate
wino-like lightest neutralino $\tilde N_1$ and chargino $\tilde C_1$. 
The next-to-lightest neutralino $\tilde N_2$ is predominantly bino because successful electroweak symmetry breaking typically requires a $\mu$-term with magnitude of a few times $M_2$.
The remaining neutralinos $\tilde N_{3,4}$ and
chargino $\tilde C_2$ are therefore heavier and Higgsino-like.

The mass scale  of the sleptons in hybrid anomaly-gravity mediation,
$\tilde m \sim \sqrt{r-g_M}  |M_{\Phi}| $, $M_{\Phi} \simeq m_{3/2}/(\lf)$,  is set by requiring no tachyons and search limits from below and by avoiding fine-tuning the Higgs mass from above.
The model predicts both left-right and flavor splitting of order one, both effects being much larger than the respective AMSB background~\cite{Gherghetta:1999sw} (2nd column).

The sfermion spectrum for the hybrid model (3rd column in Fig.~\ref{fig:spectra}) is schematic only, because a precise prediction of the average mass scale and the splittings would require a model which specifies  the values of  $r$ and at least the diagonal entries of the flavor matrices $X_{ii}^M$ more precisely than the generic anomaly-gravity framework does.
Note that the doublet masses are nondegenerate by $D$-terms, $\tilde m_{l_Li}^2-\tilde m_{\nu_L i}^2=-\cos^2\! \theta_W \cos 2 \beta \, m_Z^2$, where $\theta_W$ denotes the weak mixing angle. As already mentioned, this effect is subleading and not shown. 
It is possible that instead of a wino-like neutralino the lightest sneutrino  is the lightest supersymmetric particle, but we do not consider this any further.

In Fig.~\ref{fig:spectra} we also show the squark spectrum assuming a similar mechanism for flavor as for the leptons. The squark masses have a larger AMSB contribution and hence are more predictable than the slepton ones.
Because of the stronger RG suppression of weak scale flavor violation from the strong interaction, the mass splittings for the squarks are reduced to around ${\cal{O}}(0.1)$.

\subsection{Flavor violation at the weak scale}

For the discussion of low energy flavor and \emph{CP} violating processes it is customary to use 
mass insertion  parameters $\delta_{ij}^M$ (here in the two-generation effective framework):
\be
\delta_{ij}^M=\frac{\Delta\tilde m^2_{M ji}}{\tilde m^2}
K_{ij}^{M} K_{jj}^{M*} \, .
\ee
Here $K_{i j}^M$ is the mixing angle in the coupling of the 
bino and neutral wino to $M$-chiral leptons $l_i$ and sleptons $\tilde l_j$.
The smaller the $K_{i j}^M$, the stronger the sfermions and fermions are aligned.
Because of the order one mass splittings in the hybrid anomaly-gravity mediation model, $\Delta\tilde m^2_{M ij}/\tilde m^2 \sim {\cal{O}}(1)$, we need to rely on alignment  to control FCNC rates.

Rare decay data, most notably $l_i \to l_j \gamma$ decays, constrain the amount of  flavor
violation, that is, the $\delta_{ij}^M$. The current bounds on the branching ratios are given 
at 90 \% C.L.~as~\cite{Brooks:1999pu}
\bea
\textrm{BR}(\mu \rightarrow e \gamma) &<& 1.2 \times 10^{-11}  \, ,\nn
\\
\textrm{BR}(\tau \rightarrow e \gamma) &<& 3.3 \times 10^{-8} \, , \nn
\\
\textrm{BR}(\tau \rightarrow \mu \gamma)&<& 4.4 \times 10^{-8} \, .
\label{eq:LFVdata}
\eea
These lead to a minimum of alignment required if there is no flavor suppression present from degeneracy as in the hybrid anomaly-gravity mediation model.
The corresponding alignment bounds are given in Table \ref{tab:LFV-bounds} for generational mixing of doublets. The ones for the singlets are somewhat less severe at least for 1-2 mixing.
The flavor constraints on the mixing between the first and second generation are the strongest ones.
\begin{table}[h] 
\centering
\caption{
Required level of alignment  for order one mass splitting, $ \delta^M_{ij} \sim K_{i j}^M $, for chirality preserving slepton $i$-$j$ mixing between the $i$th and $j$th generation from rare lepton decay data, adopted from Ref.~\cite{Ciuchini:2007ha} with updates from current data Eq.~(\ref{eq:LFVdata}).
}
 \label{tab:LFV-bounds}
  \begin{tabular}{c|c|c|c}
  \hline \hline
  $i$-$j$ mixing&1-2 & 1-3 & 2-3\\
  \hline
  $\delta_{i j}^L \lesssim$ & $6 \times 10^{-4}$ &0.08& 0.10 \\
  \hline \hline 
  \end{tabular}
\end{table}

The $\delta^{M}_{ij}$ parameters can also be written as the $M$-chiral mass-squared matrix in the basis where the leptons are mass eigenstates and the neutral gaugino interactions are flavor universal, divided by an average mass-squared.
In anomaly-gravity mediation we obtain
\be
\delta^{M}_{ij} \sim \frac{r}{-g_M+r} (V^{M \dagger} X^M V^M)_{ij} \,, \quad ( i \neq j).
\ee
Here, $V^{R}, V^L$ are unitary matrices which  bring the lepton Yukawa matrix to diagonal form
as $V^{R \dag} Y^T_E V^L$.

If the diagonal entries of $X_M$ are all of the same order of magnitude and the off-diagonal entries are suppressed and, furthermore, the Yukawa matrix is hierarchical so that the diagonal elements of $V^M$ are of order unity and the off-diagonal entries are suppressed, one has (cf.~\cite{Feng:2007ke}), using $r \sim 1$,
\be  \label{eq:delta-eq}
\delta^{M}_{ij} \sim  \textrm{max} \left\lbrace |X^M_{ij}|, |V^M_{ij}|, |V^M_{ji}| \right \rbrace \,.
\ee

\subsection{Origin of flavor}

It is conceivable that  the mechanism that structures the PMSB soft terms and the one which produces SM flavor is related. 
We require the flavor model to reproduce the observed hierarchical charged lepton spectrum
\be \label{massratio}
\frac{m_e}{m_{\mu}}\sim {\cal{O}}(10^{-2}) \,, \quad
\frac{m_{\mu}}{m_{\tau}}\sim {\cal{O}}(10^{-1}) \,,
\ee
and an anarchical Maki-Nakagawa-Sakata (MNS) matrix 
\be \label{eq:VMNS}
(V_{\textrm{MNS}})_{ij} \sim {\cal{O}}(1) \, .
\ee
The current level of suppression of the 1-3 lepton mixing angle is considered accidental.

We wish to accommodate the masses of the light neutrinos, $m_{\nu i} $, within a supersymmetric seesaw mechanism. 
The natural way to obtain (\ref{eq:VMNS}) is through an anarchical Yukawa $Y_N$ and an anarchical mass matrix for the right-handed (RH) neutrinos, $(M_R)_{ij} \sim \hat M_R$. 
The mass scale of the RH neutrinos, $\hat M_R$, is constrained by leptogenesis.
In case of hierarchical RH neutrinos, the lightest one should have a mass of at least $10^7$ to $10^9\ \GeV$~\cite{Davidson:2002qv}.
For a moderate degeneracy this bound may even be significantly lower~\cite{Pilaftsis:2003gt}.

As an explicit realization, we consider a flavor symmetry which is a discrete subgroup of \mbox{$U(1)_p \times U(1)_q$}~\cite{Grossman:1998jj}.
Both U(1)'s are spontaneously broken by the VEV of a spurion with charge $-1$.
We further assume for simplicity that $\lambda_p \sim \lambda_q \sim \lambda$, where $\lambda_{p,q}$ are the ratios between the spurion VEV and the heavy messenger mass scale of the FN model.
This does not mean that the spurion VEVs need to be identical; we only assume that there is no hierarchy between them.
A reasonable value for the common expansion parameter  $ \lambda$ is $\sim 0.2$.
The  RH neutrinos as well as the Higgs fields are assumed to be neutral under the flavor symmetry.
For definiteness, we choose $y_{\tau} \sim \lambda^2$ for the Yukawa of the tau lepton, consistent with our assumption that $\tb$ is not large.

Under the above assumptions, the elements within each column of $Y_E$ are either of the same order of magnitude or vanish.
Then, it is straightforward to see that~-- in order for the off-diagonal elements of $V^M$ to have sufficient suppression so that the bounds from Table~\ref{tab:LFV-bounds} can be satisfied~-- the upper-right triangle of $Y_E$ must vanish (by holomorphy).
A realistic possibility for $Y_E$ is thus 
\bea  \label{eq:YE-desired}
Y_{E}
\sim
\lambda^2
\left( 
\begin{array}{ccc}
\lambda^5 & 0 & 0 \\
 \lambda^5 & \lambda^2 & 0 \\
  \lambda^5 & \lambda^2 & 1 
\end{array}
\right) 
\,,  
\eea
which implies
\be \label{eq:VM-pred} 
V^L_{ij} \sim \frac{m_{l_i}^2}{m_{l_j}^2} \,, \quad V^R_{ij} \sim \frac{m_{l_i}}{m_{l_j}}, ~~ (i<j)\,; \quad V^M_{ji}\sim V^M_{ij} \,.
\ee
This sets a lower limit on the $\delta^M$'s, independent of the choice of the FN charges, once we stick to the Yukawa matrix $Y_E$ given in Eq.~(\ref{eq:YE-desired}).

One can, for instance, realize Eq.~(\ref{eq:YE-desired}) by assigning the following charges to the lepton doublet (singlet) superfields $L_i (\bar E_i)$:
\begin{alignat}{3}
L_1: (3,0),\ & L_2: (1,2),\ && L_3: (0,3) \, ,
\nn \\
\bar E_1: (3,1),\ & \bar E_2: (2,-1),\ && \bar E_3: (2,-3) \, .  \label{eq:FNmodel}
\end{alignat}
This leads to
\be
\label{Xmatrices}
X^L
\sim
\left( 
\begin{array}{ccc}
 1 & \lambda^4 & \lambda^6 \\
 \lambda^4 & 1 & \lambda^2 \\
 \lambda^6 & \lambda^2 & 1 
\end{array}
\right) , \
X^R
\sim 
\left( 
\begin{array}{ccc}
  1 & \lambda^3 & \lambda^5 \\
 \lambda^3 & 1 & \lambda^2 \\
  \lambda^5 & \lambda^2 & 1 
\end{array}
\right),
\ee
and $\delta^{M}_{ij} \sim X^M_{ij}$, see Eq.~(\ref{eq:delta-eq}).

By construction, the neutrino sector is anarchical:
\bea \label{neutrinoanarchy}
(Y_{N})_{ij}&\sim&
\lambda^{n_{\nu}} , ~
(m_\nu)_{ij} \sim
\frac{\vev{H^{0}_{u}}^2}{\hat M_{R}} \lambda^{2 n_{\nu}} \,,
 \quad \forall i,j \,,
\eea
with $n_{\nu}=3$ for the model Eq.~(\ref{eq:FNmodel}).
($\vev{H^{0}_{u (d)}}$ denotes the VEV of the scalar component of the neutral Higgs superfield with hypercharge $+1/2$ $(-1/2)$.)
In order to arrive at a mass scale of the light neutrinos of around  $0.1$~eV, the RH neutrinos should have a mass scale $\hat M_{R}\sim 10^{10}\ \GeV$, compatible with  leptogenesis.

We also consider flavor violation from the seesaw sector~\cite{Hisano:1995cp}.
The dominant contribution is due to $Y_N$, which contributes to the RG-running between the high scale and the mass scale of the RH neutrinos.
This gives (for $i\neq j$)
\be
\frac{\delta X^L_{ij}}{X^L_{ij}} 
\sim 
\frac{\ln(M_*/\hat M_R)}{\lf} \frac{\lambda^{6}}{X^L_{ij}} 
\simeq 
0.1 \frac{\lambda^{6}}{X^L_{ij}} \,,
\ee
which is negligible for all off-diagonal entries.

We conclude that the FN model Eq.~(\ref{eq:FNmodel}) is a viable solution to the flavor and tachyon problem within anomaly-gravity mediation. The alignment constraints on the $\delta^M_{ij}$ parameters given in Table~\ref{tab:LFV-bounds} are fulfilled.
Note that our model is similar to model~A of Ref.~\cite{Feng:2007ke}.  
Compared to the latter, our charge assignments lead to a somewhat stronger alignment for the $1$-$j$ singlet and a lesser one for the $i$-$3$ doublet slepton mixings.

\subsection{Stronger alignment}

The model Eq.~(\ref{eq:FNmodel}) corresponds, at least with respect to 1-2 mixing, to one with a minimal amount of alignment.
Hence, the rare $\mu \to e \gamma$ decays could be just around the corner, and $\tau \to \mu \gamma$ is not far away, too.
We ask here about the consequences if future data would require a higher level of alignment.

First, recall that requiring neutrino anarchy asks for Eq.~(\ref{neutrinoanarchy}).
Note that the integer $n_\nu$ should not be too large~-- say, $n_\nu \lesssim 5$~-- otherwise the seesaw scale gets too low for leptogenesis. On the other hand, $ 2n_{\nu}$ must be at least as large as the largest exponent of $\lambda$ in $X^L$.
(This is seen as follows:
Since the neutrino sector should be anarchical, we need $p_{L_i}+q_{L_i}=n_{\nu} \quad \forall i $, and all charges of the $L_i$ denoted by $(p_{Li},q_{Li})$ are positive to avoid holomorphic zeros in $Y_N$.
This implies $0\leq q_{Li}+p_{Lj}\leq 2 n_\nu$, and one arrives at
$
\max_{ij} \{n_{X_{\tilde L}ij}\}=2 \max_{ij} \{|p_{Li}-p_{Lj}|\}=2 \max_{ij} \{|n_\nu-(q_{Li}+p_{Lj})|\} \leq 2 n_\nu \,.
$)
The upper limit on $n_\nu$ hence bounds the alignment factors $K^L_{ij}$ from below.

One could for instance consider an example with $n_\nu = 5$.
We keep $Y_E$ as in Eq.~(\ref{eq:YE-desired}) and Eq.~(\ref{eq:VM-pred}) holds.
The maximal achievable amount of alignment, $K^M_{ij}\sim V^M_{ij}$, is then obtained by having $X^M \sim V^M$.
A $U(1) \times U(1)$ flavor charge assignment which realizes this is
\begin{alignat}{3}
L_1: (5,0),\ & L_2: (2,3),\ && L_3: (0,5) \, ,
\nn \\
\bar E_1: (1,1),\ & \bar E_2: (1,-2),\ && \bar E_3: (1,-4) \, .  \label{eq:FNmodelc}
\end{alignat}

To obtain even stronger alignment, one needs one or more holomorphic zeros in the lower-left triangle of $Y_E$.
Keeping $n_\nu = 5$, we can at most obtain $X^L_{12}\sim \lambda^{10}$.
If one wants to realize $\delta^L_{12}\sim \lambda^{10}$, one should then have a suppression of $\textrm{max}  \{|V^L_{12}|, |V^L_{21}|\} $ with at least $\lambda^{10}$.
Consider for instance the charge assignment 
\begin{alignat}{4} \label{highalignment}
&L_1: (5,0),\ && L_2: (0,5),\ && L_3: (2,3) \, ,
\nn \\
&\bar E_1: (-1,3),\ && \bar E_2: (3,-4),\ && \bar E_3: (-1,-2) \,.
\end{alignat}
This yields
\bea  
Y_{E}
\sim
\lambda^2
\left( 
\begin{array}{ccc}
\lambda^5 & 0 & 0 \\
0 & \lambda^2 & 0 \\
  \lambda^5 & 0 & 1 
\end{array}
\right)  
\eea
--~in which case the only nonvanishing off-diagonal elements of the $V^M$ are
$
V^L_{13}  \sim V^L_{31} \sim m_{e}^2/m_{\tau}^2 \,,  V^R_{13} \sim V^R_{31} \sim m_{e}/m_{\tau}
$~--
and
\be
X^L
\sim
\left( 
\begin{array}{ccc}
 1 & \lambda^{10} & \lambda^6 \\
 \lambda^{10} & 1 & \lambda^4 \\
 \lambda^6 & \lambda^4 & 1 
\end{array}
\right) , 
X^R
\sim 
\left( 
\begin{array}{ccc}
  1 & \lambda^{11} & \lambda^5 \\
 \lambda^{11} & 1 & \lambda^6 \\
  \lambda^5 & \lambda^6 & 1 
\end{array}
\right) .
\ee
We obtain suppressed $\delta^M_{ij} \sim X^M_{ij}$, which is far beyond the reach of future experiments~  \cite{Signorelli:2003vw,Bona:2007qt}.

In summary, we find that, while the possible alignment is not unlimited in $U(1) \times U(1)$ FN models, this limit is weak enough such that bounds from upcoming FCNC tests can be evaded; see the last example above.

\subsection{Flavor from wavefunctions}

Flavored wavefunctions are known to provide a significant amount of flavor alignment, 
in particular if the trilinear soft terms are small~\cite{Davidson:2007si,Nomura:2007ap}. To accommodate the lepton spectrum 
Eq.~(\ref{massratio}) and mixing Eq.~(\ref{eq:VMNS}) one finds
\be \label{eq:WFRratio}
\frac{X^L_{ii}}{X^L_{jj}}\sim {\cal{O}}(1) \,, \quad
\frac{X^R_{ii}}{X^R_{jj}}\sim \frac{m^2_{l_i}}{m^2_{l_j}}  \,.
\ee
The doublets have $X_{ii}^L$ of similar size in order to accommodate order-one $V_{\mathrm{MNS}}$ entries.
However, the $X_{ii}^R$ are strongly hierarchical and decrease towards the lighter lepton generations. They will not be large enough to remove all the tachyons while keeping the stau mass of electroweak size. It follows that the mechanism of wavefunction renormalization does not naturally work  within anomaly-gravity mediation.

\section{Phenomenology \label{sec:pred}}

We predict the generic flavor-blind features of AMSB, with an interesting flavor phenomenology
driven by flavor alignment. This includes:\\

{\it Lepton FCNC decays:} 
Working flavor models, such as the one in Eq.~(\ref{eq:FNmodel}), often predict $\mu \to e \gamma$ rates close to the current experimental limit given in Eq.~(\ref{eq:LFVdata}).
The MEG collaboration expects to improve its  reach in the $\mu \to e \gamma$ branching ratio by about 2 orders of magnitude in the next few years~\cite{Signorelli:2003vw}. 
This will allow to access even more aligned models.
In some cases, as in the model Eq.~(\ref{eq:FNmodel}), the prediction for $\tau \to \mu \gamma$ is close to the present data as well and could be probed at a possible super flavor factory with reach down to $2 \times 10^{-9}$~\cite{Bona:2007qt} (with $75\ {\rm ab}^{-1}$).

However, the framework presented allows for the possibility that the FCNC decays escape the detection of even the future planned searches, see the model Eq.~(\ref{highalignment}).\\

{\it Slepton spectroscopy at colliders:}  
The collider phenomenology is different from the common MSSM variants  because of the large mass splitting between singlet and doublet sleptons and between different generations; see Eq.~(\ref{eq:masssplit}).
This leads to six distinct charged sleptons with electroweak-scale  masses, see Fig.~\ref{fig:spectra}.

The splittings can be accessed  from dilepton spectra with missing energy in $\tilde N_2 \to l \tilde l^*,\bar  l \tilde l \to \bar l l \tilde N_1$ cascades.
The latter are sensitive to the mass of the intermediate slepton.
Because of the high level of alignment in the hybrid model, lepton flavor violating modes are strongly suppressed, and we consider only same flavor dileptons.

The selectron-smuon mass difference can be probed by comparing dimuon versus dielectron spectra, and their respective kinematical edges.
Both spectra should exhibit a multi-edge structure from $\tilde l_R$ and $\tilde l_L$ exchanges.
The features of the hybrid spectrum suggest that there is no strong hierarchy between the distributions of both chiralities and all flavors:
The $\tilde N_2$ is mostly gaugino-bino, and the sleptons are produced flavor-universally via $\tilde N_2 \to l_{iM} \tilde l^*_{iM} , \bar l_{iM} \tilde l_{iM}$.
Because of the different hypercharges, the branching ratio into singlets is 4 times larger than the one into doublets modulo phase space effects.
Subsequently, the sleptons decay to  the ${\tilde N_1}$-lightest supersymmetric particle plus $l_{iM}$ or $\bar l_{iM}$.
Because of the mostly wino-nature of the  ${\tilde N_1}$ the latter decays will be more rapid for the doublet sleptons, but the branching ratios again will be roughly in the same ballpark barring phase space cancellations.

LHC studies  to measure the selectron-smuon mass difference exist for mSUGRA scenarios~\cite{Allanach:2008ib,CMS}.
The expected sensitivities  are very promising, and reach down to mass splittings much smaller than the order one prediction of the hybrid model.
Depending on the SUSY scale, measurements can already be performed with rather moderate luminosities~\cite{CMS}.
Because of the different  AMSB-like gaugino pattern one would need to perform a dedicated analysis for anomaly-gravity mediation to quantify its LHC prospects, which is beyond the scope of this work.

Note that a measurement of the selectron doublet-singlet splitting is a prime application for an $e^+ e^-$ high energy linear collider~\cite{AguilarSaavedra:2001rg}. \\

{\it Lepton electric dipole moments (EDMs):} 
If flavor violation comes along with \emph{CP} phases it is possible to see this in the muon EDM $d_{\mu}$.
Using the same approximations as in~\cite{Hiller:2010ib}, one roughly has
\be
|d_{\mu}| \sim {\rm Im}(\delta_{23}^L \delta_{23}^{R*})   \frac{\tb\, m_{\tau}}{\tilde m} 10^{-20}  \,  {\rm e cm} \, .
\ee
The factor proportional to the tau mass provides the requisite left-right mixing of the stau in case of a heavy $\mu$-term as in AMSB.
In the model  Eq.~(\ref{eq:FNmodel}), $\delta^{L}_{23}\sim \delta^{R}_{23}\sim \lambda^2$ holds, and muon EDMs from slepton flavor with \emph{CP} violation up to order $10^{-24} \, {\rm e cm}$ are possible.
This is well below the current bound $d_{\mu}=(-0.1\pm0.9) \times 10^{-19} \, {\rm e cm}$~\cite{Bennett:2008dy}, but within reach of the proposed activities to measure the muon EDM as low as $5 \times 10^{-25} \, {\rm e cm}$~\cite{Adelmann:2006ab}. 
The electron EDM has a stronger experimental bound, but at the same time, the
$\delta_{13}^M$ parameters are suppressed even further, so that the sensitivity for flavored \emph{CP} violation is larger in $d_\mu$.\\

{\it The quark sector: } 
As flavor symmetries between lepton and quarks may or may not be related, we can make a generic alignment prediction only: $\Ocal(10\%)$ squark mass-squared splittings.
The latter is smaller than for sleptons due to the large flavor-universal RG-effect of the gluinos, which is absent for sleptons. 
Known predictions include $D$-$\bar D$ mixing close to the experimental limit~\cite{Feng:2007ke} and a lower bound for hadron EDMs~\cite{Altmannshofer:2010ad}.\\

\section{Conclusions} \label{sec:conclusions}

The forthcoming explorations of the TeV-scale might reveal patterns which do not fit into the expectations based on standard models beyond the SM. 
Models with mixed sources of SUSY breaking can provide interesting alternatives.
The motivation to investigate hybrid models goes, however, much further: 
They offer new insights into model space and problems and are often not unnatural from a model-building perspective.

The framework of  hybrid anomaly-gravity mediation has several virtues regarding model building and phenomenology:
the tachyons of anomaly mediation are removed  by flavorful contributions, and the low energy spectrum contains imprints of the origin of flavor symmetry breaking. 
The gravitino can be heavy enough that it decays before nucleosynthesis. 
The gaugino masses are AMSB-like, with known LHC signatures (see {\it e.g.}~\cite{Gherghetta:1999sw}). 
The hybrid model exhibits a very characteristic slepton spectrum with order one mass splittings between different generations and between left- and right-handed sleptons.

We suggest to pursue LHC dilepton searches  with missing energy in both dielectron and dimuon final states. 
If accessible, the corresponding tau spectra are similarly informative.
The contributions from sleptons of both chiralities induce a multi-edge structure. 
Because of the high level of alignment required, opportunities exist for lepton flavor precision searches for FCNC decays and, if \emph{CP} violation is linked to flavor violation, the muon EDM.

\begin{acknowledgements}
We thank A.~Hebecker and M.~Schmaltz for useful comments on an earlier version of the manuscript and J.~S.~Kim for an enlightening discussion about slepton spectroscopy at colliders. C.~G. thanks E.~Pajer for correspondence. This work is supported in part by the {\it Bundesministerium f\"ur Bildung und Forschung (BMBF)}.
\end{acknowledgements}



\begin{thebibliography}{}

\bibitem{Isidori:2010kg}
  G.~Isidori, Y.~Nir and G.~Perez,
  arXiv:1002.0900 [hep-ph].
  
\bibitem{Randall:1998uk}
 L.~Randall and R.~Sundrum,
  Nucl. Phys.\  B\ {\bf 557} (1999) 79.

\bibitem{Giudice:1998xp}
  G.~F.~Giudice, M.~A.~Luty, H.~Murayama and R.~Rattazzi,
  JHEP {\bf 9812} (1998) 027.
  
\bibitem{Allanach:2009ne}
  B.~C.~Allanach, G.~Hiller, D.~R.~T.~Jones and P.~Slavich,
  JHEP {\bf 0904}, 088 (2009).
  
\bibitem{Pomarol:1999ie}
  A.~Pomarol and R.~Rattazzi,
  JHEP {\bf 9905} (1999) 013;
  %
  Z.~Chacko, M.~A.~Luty, I.~Maksymyk and E.~Ponton,
  JHEP {\bf 0004} (2000) 001;
  %
  E.~Katz, Y.~Shadmi and Y.~Shirman,
  JHEP {\bf 9908} (1999) 015;
  %
  I.~Jack and D.~R.~T.~Jones,
  Phys.\ Lett.\  B {\bf 482} (2000) 167;
  %
  M.~S.~Carena, K.~Huitu and T.~Kobayashi,
  Nucl.\ Phys.\  B {\bf 592} (2001) 164;
  %
  B.~C.~Allanach and A.~Dedes,
  JHEP {\bf 0006} (2000) 017;
  %
  D.~E.~Kaplan and G.~D.~Kribs,
  JHEP {\bf 0009} (2000) 048;
%
  N.~Arkani-Hamed, D.~E.~Kaplan, H.~Murayama and Y.~Nomura,
  JHEP {\bf 0102} (2001) 041;
  %
  Z.~Chacko and M.~A.~Luty,
  JHEP {\bf 0205} (2002) 047;
  %
  A.~E.~Nelson and N.~J.~Weiner,
  Phys.\ Rev.\ Lett.\  {\bf 88} (2002) 231802;
  %
  N.~Okada,
  Phys.\ Rev.\  D {\bf 65} (2002) 115009;
%
I.~Jack and D.~R.~T.~Jones,
Nucl.\ Phys.\ B {\bf 662} (2003) 63;
  %
  O.~C.~Anoka, K.~S.~Babu and I.~Gogoladze,
  Nucl.\ Phys.\  B {\bf 686} (2004) 135;
  %
  R.~Kitano, G.~D.~Kribs and H.~Murayama,
  Phys.\ Rev.\  D {\bf 70} (2004) 035001;
  %
  M.~Ibe, R.~Kitano, H.~Murayama and T.~Yanagida,
  Phys.\ Rev.\  D {\bf 70} (2004) 075012;
    %
  R.~Sundrum,
  Phys.\ Rev.\  D {\bf 71} (2005) 085003;
  %
  M.~Ibe, R.~Kitano and H.~Murayama,
  Phys.\ Rev.\  D {\bf 71} (2005) 075003;
  %
  R.~N.~Mohapatra, N.~Setzer and S.~Spinner,
  JHEP {\bf 0804} (2008) 091;
  %
  J.~de Blas, P.~Langacker, G.~Paz and L.~T.~Wang,
  JHEP {\bf 1001} (2010) 037.
  
\bibitem{Gherghetta:1999sw}
  T.~Gherghetta, G.~F.~Giudice and J.~D.~Wells,
  Nucl.\ Phys.\  B {\bf 559} (1999) 27.

\bibitem{Gregoire:2005jr}
  T.~Gregoire, R.~Rattazzi and C.~A.~Scrucca,
  Phys.\ Lett.\  B {\bf 624} (2005) 260.
  
\bibitem{Feng:2007ke}
J.~L.~Feng, C.~G.~Lester, Y.~Nir and Y.~Shadmi,
Phys.\ Rev.\ D {\bf 77} (2008) 076002.

\bibitem{Nomura:2007ap}
  Y.~Nomura, M.~Papucci and D.~Stolarski,
  Phys.\ Rev.\  D {\bf 77}, 075006 (2008).

\bibitem{Hiller:2010dv}
  G.~Hiller, Y.~Hochberg, Y.~Nir,
  JHEP {\bf 1003}, 079 (2010).
  
\bibitem{Nir:1993mx}
Y.~Nir and N.~Seiberg,
Phys.\ Lett.\ B {\bf 309} (1993) 337.

\bibitem{Luty:2001jh}
  M.~A.~Luty and R.~Sundrum,
  Phys.\ Rev.\  D {\bf 65} (2002) 066004.
 
\bibitem{Anisimov:2001zz}
  A.~Anisimov, M.~Dine, M.~Graesser and S.~D.~Thomas,
  Phys.\ Rev.\  D {\bf 65} (2002) 105011;
  %
  M.~Dine, P.~J.~Fox, E.~Gorbatov, Y.~Shadmi, Y.~Shirman and S.~D.~Thomas,
  Phys.\ Rev.\  D {\bf 70} (2004) 045023;
  %
  M.~Berg, D.~Marsh, L.~McAllister and E.~Pajer,
  arXiv:1012.1858 [hep-th].
   
\bibitem{Kaplan:1999ac}
  D.~E.~Kaplan, G.~D.~Kribs and M.~Schmaltz,
  Phys.\ Rev.\  D {\bf 62} (2000) 035010;
  %
  Z.~Chacko, M.~A.~Luty, A.~E.~Nelson and E.~Ponton,
  JHEP {\bf 0001} (2000) 003.
  
\bibitem{Chacko:2000fn}
  Z.~Chacko and M.~A.~Luty,
  JHEP {\bf 0105} (2001) 067.
  
\bibitem{Choi:2004sx}
  K.~Choi, A.~Falkowski, H.~P.~Nilles, M.~Olechowski and S.~Pokorski,
  JHEP {\bf 0411} (2004) 076;
  %
  K.~Choi, A.~Falkowski, H.~P.~Nilles and M.~Olechowski,
  Nucl.\ Phys.\  B {\bf 718} (2005) 113;
  %
  K.~Choi, K.~S.~Jeong and K.~i.~Okumura,
  JHEP {\bf 0509} (2005) 039.
  
\bibitem{Buchbinder:2003qu}
  I.~L.~Buchbinder, S.~J.~J.~Gates, H.~S.~J.~Goh, W.~D.~I.~Linch, M.~A.~Luty, S.~P.~Ng and J.~Phillips,
  Phys.\ Rev.\  D {\bf 70} (2004) 025008;
  %
  R.~Rattazzi, C.~A.~Scrucca and A.~Strumia,
  Nucl.\ Phys.\  B {\bf 674} (2003) 171.
  
\bibitem{Luty:1999cz}
  M.~A.~Luty and R.~Sundrum,
  Phys.\ Rev.\  D {\bf 62} (2000) 035008.

\bibitem{Witten:1996mz}
  P.~Horava and E.~Witten,
  Nucl.\ Phys.\  B {\bf 460} (1996) 506;
%
  E.~Witten,
  Nucl.\ Phys.\  B {\bf 471} (1996) 135.

\bibitem{Kachru:2006em}
  S.~Kachru, J.~McGreevy and P.~Svrcek,
  JHEP {\bf 0604} (2006) 023.
 
\bibitem{Paige:2003mg}
  F.~E.~Paige, S.~D.~Protopopescu, H.~Baer and X.~Tata,
  arXiv:hep-ph/0312045.
  
\bibitem{Brooks:1999pu}
M.~L.~Brooks {\it et al.}[MEGA Collaboration],
Phys.\ Rev.\ Lett.\ {\bf 83} (1999) 1521.
%
B.~Aubert {\it et al.}[BABAR Collaboration],
Phys.\ Rev.\ Lett.\ {\bf 104} (2010) 021802.
  
\bibitem{Ciuchini:2007ha}
  M.~Ciuchini, A.~Masiero, P.~Paradisi {\it et al.},
  Nucl.\ Phys.\  {\bf B783}, 112-142 (2007).
    
\bibitem{Davidson:2002qv}
  W.~Buchm\"uller and M.~Pl\"umacher,
  Int.\ J.\ Mod.\ Phys.\  A {\bf 15} (2000) 5047;
  S.~Davidson and A.~Ibarra,
  Phys.\ Lett.\  B {\bf 535} (2002) 25;
%
  G.~F.~Giudice, A.~Notari, M.~Raidal, A.~Riotto and A.~Strumia,
  Nucl.\ Phys.\  B {\bf 685} (2004) 89.

  
\bibitem{Pilaftsis:2003gt}
  A.~Pilaftsis and T.~E.~J.~Underwood,
  Nucl.\ Phys.\  B {\bf 692} (2004) 303;
  %
  T.~Hambye, Y.~Lin, A.~Notari, M.~Papucci and A.~Strumia,
  Nucl.\ Phys.\  B {\bf 695} (2004) 169.
  
\bibitem{Grossman:1998jj}
  Y.~Grossman, Y.~Nir, Y.~Shadmi,
  JHEP {\bf 9810}, 007 (1998).
  
\bibitem{Hisano:1995cp}
  F.~Borzumati and A.~Masiero,
  Phys.\ Rev.\ Lett.\  {\bf 57} (1986) 961;
  %
  J.~Hisano, T.~Moroi, K.~Tobe and M.~Yamaguchi,
  Phys.\ Rev.\  D {\bf 53}, 2442 (1996).

\bibitem{Signorelli:2003vw}
  G.~Signorelli,
  J.\ Phys.\ G {\bf 29}, 2027 (2003).

\bibitem{Bona:2007qt}
  M.~Bona {\it et al.},
  arXiv:0709.0451 [hep-ex].

\bibitem{Davidson:2007si}
  S.~Davidson, G.~Isidori and S.~Uhlig,
  Phys.\ Lett.\  B {\bf 663}, 73 (2008);
  %
  E.~Dudas, G.~von Gersdorff, J.~Parmentier and S.~Pokorski,
  JHEP {\bf 1012}, 015 (2010).
  
\bibitem{Allanach:2008ib}
 B.~C.~Allanach, J.~P.~Conlon, C.~G.~Lester,
 Phys.\ Rev.\  {\bf D77}, 076006 (2008);
%
  A.~J.~Buras, L.~Calibbi and P.~Paradisi,
  JHEP {\bf 1006} (2010) 042;
 A.~Abada, A.~J.~R.~Figueiredo, J.~C.~Romao and A.~M.~Teixeira,
 JHEP {\bf 1010}, 104 (2010).

\bibitem{CMS} [CMS Collaboration] CMS PAS SUS-09-002.

\bibitem{AguilarSaavedra:2001rg}
  J.~A.~Aguilar-Saavedra {\it et al.}  [ECFA/DESY LC Physics Working Group],
  arXiv:hep-ph/0106315.

\bibitem{Hiller:2010ib}
  G.~Hiller, K.~Huitu, T.~Ruppell and J.~Laamanen,
  Phys.\ Rev.\  D {\bf 82} (2010) 093015.

\bibitem{Bennett:2008dy}
  G.~W.~Bennett {\it et al.}  [Muon (g-2) Collaboration],
  Phys.\ Rev.\  D {\bf 80} (2009) 052008.

\bibitem{Adelmann:2006ab}
  A.~Adelmann {\it et al.},
  arXiv:hep-ex/0606034v2;
  A.~Adelmann, K.~Kirch, C.~J.~G.~Onderwater and T.~Schietinger,
  J.\ Phys.\ G {\bf 37}, 085001 (2010).
  
\bibitem{Altmannshofer:2010ad}
  W.~Altmannshofer, A.~J.~Buras and P.~Paradisi,
  Phys.\ Lett.\  B {\bf 688}, 202 (2010).
 
\end{thebibliography}
\end{document}